\documentclass[10pt]{article}
\usepackage{OSAmeet}
\def\ket#1{| #1 \rangle}
\def\bra#1{\langle #1 |}

\begin{document}

\title{Robust quantum memory via quantum control}
\HeaderAuthorTitleMtg{Greentree, Schirmer and Solomon}{Robust quantum memory via quantum control}{ICQI}
\author{Andrew D. Greentree, S.~G.\ Schirmer and A.~I.\ Solomon}
\address{Quantum Processes Group, Departments of Physics and Astronomy and Applied Maths,
         The Open University, Milton Keynes, MK7 6AA, United Kingdom}
\email{Tel: +44-1908-652326, Fax: +44-1908-652140, S.G.Schirmer@open.ac.uk}

\begin{abstract}
A general scheme for building a quantum memory by transferring quantum information to 
an essentially decoherence-free memory transition using quantum control is presented
and illustrated by computer simulations.
\end{abstract}

\noindent                      
Quantum computation \cite{QC} has been a fruitful area of research lately.  Some of 
the most promising schemes involve encoding qubits into ions and neutral atoms in 
high Q optical cavities.  One of the greatest limitations to such schemes is the 
decoherence which occurs at the optical transitions.  This decoherence is the limiting
factor in determining the temporal length of a sequence of pulses to perform a given
computation.  Many schemes have been suggested to overcome this decoherence, especially
quantum error correction \cite{QCErrorCorr}, which uses redundant information to 
compensate for the losses, and decoherence-free substates \cite{DF}, which use 
combinations of states which are robust against decay because of quantum interference.
In this paper we employ Lie group decompositions \cite{Decomposition} to derive a 
promising scheme for a quantum memory, with the idea being to transfer quantum 
information from the important channel for quantum information processing to a 
``memory transition'' which holds the quantum information.  

\begin{figure}
\begin{minipage}{2.5in}
\centerline{\scalebox{0.8}{\includegraphics{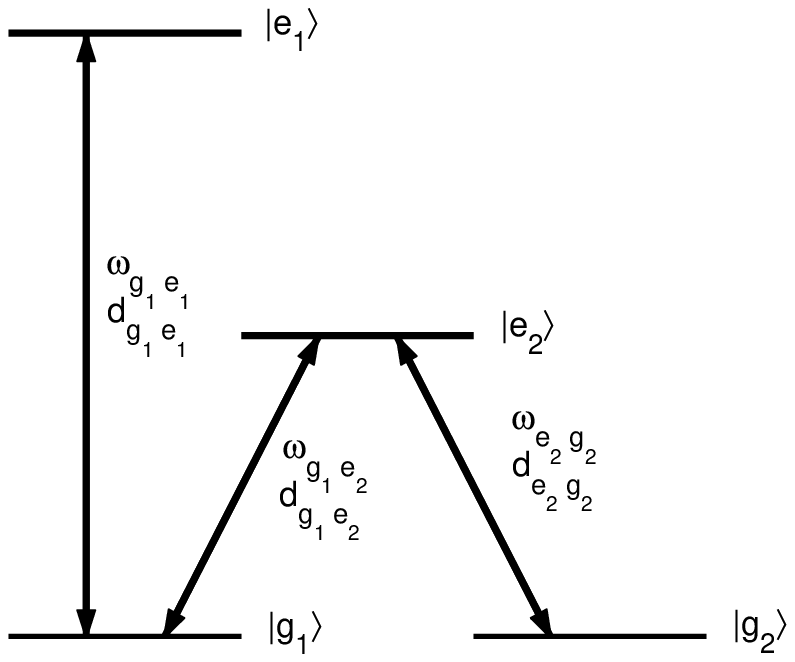}}}
\caption{Energy level diagram}\label{fig:sys}
\end{minipage}
\hfill
\begin{minipage}{4in} 
\centerline{\scalebox{0.65}{\includegraphics{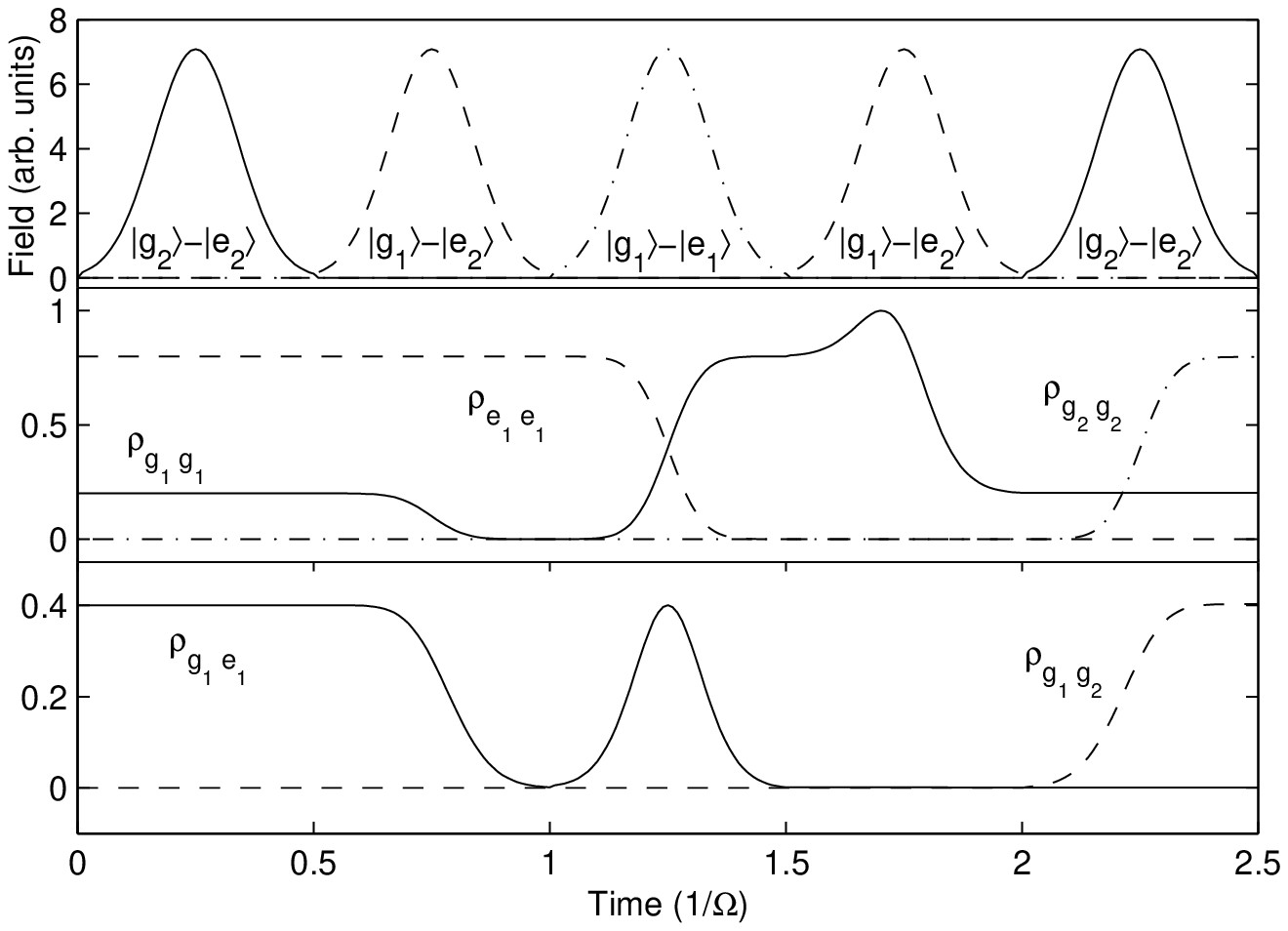}}}
\caption{Control pulses and evolution of the populations and coherence}\label{fig:rho}
\end{minipage}
\end{figure}

Our scheme can be illustrated with a simple example.  Consider a four-level atom as 
depicted in figure \ref{fig:sys} with two degenerate ground states $\ket{g_1}$ and 
$\ket{g_2}$ and two non-degenerate excited states $\ket{e_1}$ and $\ket{e_2}$.  The 
ground states might be for example Zeeman sublevels of an atom such as Rubidium,
although our scheme is far more general than this.  We assume that the $\ket{g_1}-
\ket{e_1}$ transition is an optical transition that forms part of a quantum information
processing scheme.  After performing some quantum logical operations we wish to protect
the quantum information stored in this transition in a long-lived system by applying 
a series of Gaussian control pulses (derived from optical fields) to transfer the quantum 
information onto the decoherence-free $\ket{g_1}-\ket{g_2}$ transition.  Although the
transition pairs $\ket{g_1}-\ket{e_1}$, $\ket{g_2}-\ket{e_1}$ and $\ket{g_1}-\ket{e_2}$,
$\ket{g_2}-\ket{e_2}$ are at equal frequencies, they can be individually addressed
through the choice of appropriate field polarisations, allowing complete controllability
of the system.

Formally, the problem can be stated as follows.  We wish to map the density matrix 
$\rho$ representing the (initial) state of the system, whose elements are 
$\rho_{g_1g_1}$, $\rho_{e_1e_1}$, $\rho_{g_1e_1}$ and $\rho_{g_2g_2}=\rho_{e_2e_2}=
\rho_{g_1g_2}=\rho_{g_1e_2}=\rho_{g_2 e_1}=\rho_{g_2e_2}=0$, onto a density matrix 
$\rho'$ such that $\rho_{g_1g_1}'=\rho_{g_1g_1}$, $\rho_{g_2g_2}'=\rho_{e_1e_1}$, 
$\rho_{g_1g_2}'=\rho_{g_1e_1}$ and $\rho_{e_1e_1}=\rho_{e_2e_2}=\rho_{g_1e_1}=
\rho_{g_1e_2}=\rho_{g_2e_1}=\rho_{g_2e_2}=0$ by applying a sequence of simple control
pulses.  Because the populations and coherence have been mapped to degenerate energy 
levels with no allowed transitions between them, these states will be extremely long 
lived.  The quantum information can then be returned to the optical transition simply
by using the reverse of the quantum control scheme.  In order to realize the mapping 
of $\rho$ onto $\rho'$ we find a unitary operator $U$ such that $\rho'=U\rho U^\dagger$
and decompose the operator $U$ into a product of simple unitary operators, each realized
dynamically by applying a control field 1 (2) (3) that drives the $\ket{g_2}-\ket{e_2}$
($\ket{g_1}-\ket{e_2}$) ($\ket{g_1}-\ket{e_1}$) transition, respectively.  Concretely, 
note that we have 
\begin{equation}
 \underbrace{\left(\begin{array}{cccc}
	\rho _{g_{1}g_{1}} & \rho _{g_{1}e_{1}} & 0 & 0 \\ 
	\rho _{e_{1}g_{1}}^* & \rho _{g_{2}g_{2}} & 0 & 0 \\ 
	0 & 0 & 0 & 0 \\ 
	0 & 0 & 0 & 0
 \end{array}\right)}_{\rho'}
 = \underbrace{\left( \begin{array}{cccc}
	-1 & 0 &  0 &  0 \\
         0 & 0 & -1 &  0 \\
         0 & 1 &  0 &  0 \\
         0 & 0 &  0 & -1
        \end{array} \right)}_{U}
 \underbrace{\left(\begin{array}{cccc}
	\rho _{g_{1}g_{1}} & 0 & \rho _{g_{1}e_{1}} & 0 \\ 
	0 & 0 & 0 & 0 \\ 
	\rho _{e_{1}g_{1}}^* & 0 & \rho _{e_{1}e_{1}} & 0 \\ 
	0 & 0 & 0 & 0
	\end{array}\right)}_{\rho}
 \underbrace{\left( \begin{array}{cccc}
	-1 &  0 & 0 &  0 \\
         0 &  0 & 1 &  0 \\
         0 & -1 & 0 &  0 \\
         0 &  0 & 0 & -1
        \end{array} \right)}_{U^\dagger}.
\end{equation}
Furthermore, we can express $U$ as a product $U=V_1 V_2 V_3 V_2 V_1$ where 
\begin{equation}
 V_1 = \left( \begin{array}{cccc}
	1 &  0 & 0 & 0 \\
        0 &  0 & 0 & 1 \\
        0 &  0 & 1 & 0 \\
        0 & -1 & 0 & 0 
       \end{array}\right), \quad
 V_2 = \left( \begin{array}{cccc}
	0 & 0 & 0 & 1 \\
        0 & 1 & 0 & 0 \\
        0 & 0 & 1 & 0 \\
       -1 & 0 & 0 & 0 
       \end{array}\right), \quad
 V_3 = \left( \begin{array}{cccc}
	 0 & 0 & 1 & 0 \\
         0 & 1 & 0 & 0 \\
        -1 & 0 & 0 & 0 \\
         0 & 0 & 0 & 1 
       \end{array}\right).
\end{equation}

Observing that \cite{Decomposition}
\begin{equation}
 V_1 =\exp\left[\frac{\pi}{2}\left(\ket{g_2}\bra{e_2}-\ket{e_2}\bra{g_2}\right)\right]
\end{equation}
we see that the operator $V_1$ can be dynamically realized by applying a pulse $f_1(t)
=A_1(t) e^{i\omega_{g_2 e_2}+\pi/2}$ with total pulse area $\int |A_1(t)|\,dt = 
\pi/(2 d_{g_2 e_2})$ and appropriate polarisation, where $\omega_{g_2 e_2}$ is the 
frequency and $d_{g_2 e_2}$ the absorption oscillator strength of the 
$\ket{g_2}-\ket{e_2}$ transition.  Similarly,
\begin{equation}
 V_2 = \exp\left[\frac{\pi}{2}\left(\ket{g_1}\bra{e_2}-\ket{e_2}\bra{g_1}\right)\right],
 \quad
 V_3 = \exp\left[\frac{\pi}{2}\left(\ket{g_1}\bra{e_1}-\ket{e_1}\bra{g_1}\right)\right]
\end{equation}
shows that $V_2$ [$V_3$] can be dynamically realized by applying a pulse 
$f_2(t) = A_2(t) e^{i\omega_{g_1 e_2}+\pi/2}$ 
[$f_3(t) = A_3(t) e^{i\omega_{g_1 e_1}+\pi/2}$]
with appropriate polarisation and total pulse area 
$\int |A_2(t)| \, dt = \pi/(2 d_{g_1 e_2})$
[$\int |A_3(t)| \, dt = \pi/(2 d_{g_1 e_1})$] where $\omega_{g_1 e_2}$ 
[$\omega_{g_1 e_1}$] is the frequency and $d_{g_1 e_2}$ [$d_{g_1 e_1}$]
the absorption oscillator strength of the $\ket{g_1}-\ket{e_2}$ 
[$\ket{g_1}-\ket{e_1}$] transition. 

Hence, generation of $U$ requires a sequence of five pulses, where the first pulse 
drives the transition $\ket{g_2}-\ket{e_2}$ and has pulse area $\pi/(2d_{g_2 e_2})$;
the second pulse drives the transition $\ket{g_1}-\ket{e_2}$ and has pulse area 
$\pi/(2d_{g_1 e_2})$; the third pulse drives the transition $\ket{g_1}-\ket{e_1}$ and
has pulse area $\pi/(2d_{g_1 e_1})$; the fourth pulses drives again the transition
$\ket{g_1}-\ket{e_2}$ and has pulse area $\pi/(2d_{g_1 e_2})$ and the last pulse drives
again the transition $\ket{g_2}-\ket{e_2}$ and has pulse area $\pi/(2d_{g_2 e_2})$.
Observe that only the total area of each pulse matters and hence the precise pulse
envelopes can be adjusted to suit experimental constraints.

Figure \ref{fig:rho} shows the pulse sequence and evolution of a system initially in 
the superposition state $(\ket{g_1}+2\ket{e_1})/\sqrt{5}$.  Note that initially 
$\rho_{g_1 g_1}=0.2$, $\rho_{e_1 e_1}=0.8$, $\rho_{g_1 e_1}=0.4$ and all other 
matrix elements of $\rho$ are zero.  Observe that at the final time we have indeed
$\rho_{g_1 g_1}'=0.2$, $\rho_{g_2 g_2}'=0.8$, $\rho_{g_1 g_2}=0.4$ and all other 
elements of $\rho'$ are zero.  The time unit in all plots is $1/\Omega$, where 
$\Omega$ is the Rabi frequency of the transition $\ket{g_1}-\ket{e_1}$.

One of the authors (ADG) would like to acknowledge the financial support of
the EPSRC.

\end{document}